\documentclass[12pt]{iopart}
\pdfoutput=1

\newcommand{\eq}[1]{Eq.\,(\ref{#1})}  
\newcommand{\eqs}[2]{Eqs.\,(\ref{#1}) and (\ref{#2})}  
\newcommand{\fig}[1]{Fig.\,\ref{#1}}  
\newcommand{\Fig}[1]{Figure \ref{#1}}  
\newcommand{\be}[1]{\begin{equation}\label{#1}}  
\newcommand{\ee}{\end{equation}}  
\newcommand{\void}[1]{}
\renewcommand{\vec}[1]{\mathbf #1}

\def\vee{v_\mathrm{ee}}
\def\vc{v_\mathrm{c}}
\def\Nph{N_\mathrm{ph}}
\def\sec#1{Sect.\,\ref{#1}}  

\usepackage{graphicx}
\usepackage{citesort}

\begin{document}
\title{Dynamics of photo-activated Coulomb complexes}

\author{Christian Gnodtke$^1$, Ulf Saalmann$^{1,2}$, Jan-Michael Rost$^{1,2}$} 
\address{$^1$Max Planck Institute for the Physics of Complex Systems\\ 
N\"othnitzer Stra{\ss}e 38, 01187 Dresden, Germany\\
$^2$Max Planck Advanced Study Group at the CFEL\\
 Luruper Chaussee 149, 22761 Hamburg, Germany} 
\date{today}
\ead{rost@pks.mpg.de}
\begin{abstract}
Intense light with frequencies above typical atomic or molecular ionization potentials as provided by free-electron lasers couples many photons into extended targets such as clusters and biomolecules. This implies, in contrast to traditional multi-photon ionization, multiple single-photon absorption.  Thereby, many electrons are removed from their bound states and either released or trapped if the target charge has become sufficiently large. We develop a simple model for this photo activation to study electron migration and interaction. It satisfies scaling relations which help to relate quite different scenarios. To understand this type of multi-electron dynamics on very short time scales is vital for assessing the radiation damage inflicted by that type of radiation and to pave the way for coherent diffraction imaging of single molecules.
\end{abstract}


\maketitle

\section{Introduction}\label{sec:intro}
The difference between multi-photon ionization at near-infrared frequencies ($\sim$\,1.5\,eV) and at extreme-ultraviolet (XUV) frequencies or higher ($>$\,50\,eV) lies in the fact that multi-photon \emph{single}-electron ionization is drastically suppressed in the latter case due to the small dipole matrix elements for continuum-continuum transitions which typically would be required for an electron to absorb more than one photon.
 As a consequence, multi-photon ionization by XUV or X-ray radiation, as available from novel free-electron laser sources \cite{acas+07,shta+10,emak+10}, means \emph{single}-photon ionization of \emph{many} atoms in an extended target such as a cluster or bio-molecule \cite{fuli+09,both+10,newo+00}. We call this process photo activation.
As other absorption mechanisms, such as inverse bremsstrahlung, are less important for high laser frequencies, the system is ``driven'' by this activation process only.  
In the following we will develop a fairly general and simple model, which we call Coulomb complex (CC). 
It focuses on the multi-electron dynamics  treating the ions created as a homogeneous and static background charge.
 The model shares similarities with the shell model for nuclei or metal clusters \cite{br93} with the difference that the electrons are through photo activation far above the ground-state which allows us to follow their dynamics classically.
 
We will first discuss the initial multi-electron ground state in \sec{sec:initial}. The ground state is the prerequisite for photo activation which we specify in \sec{sec:activation} including the set-up of a universal time-dependent photo activation rate. In \sec{sec:scaling} we discuss a powerful scaling property of the photo-activated CC which allows one to relate quite different 
 situations in terms of photo activation time (length of the XUV pulse) $T$, excess energy $\varepsilon^*$ of the electrons after photo absorption and size of the target designated by a radius $R$, to each other.
In \sec{sec:tdspectrum} we elaborate on a generic example and discuss relations of the present model to recently measured electron spectra of Xenon clusters \cite{both+10} exposed to 90\,eV photon energy pulses at FLASH \cite{acas+07}.
In \ref{sec:sequential} we give an analytical expression for the electron spectrum for the case of sequential ionization. 

\section{The Coulomb complex before photo activation}\label{sec:initial}

We assume a system with $N$ electrons which will be eventually photo-activated. They are bound by positive ions which are assumed to be fixed in space during the relevant electron dynamics. Moreover, we approximate the ions  within the CC as a homogeneously charged sphere, with total charge $Q$ and radius $R$, producing a radial potential for the electrons of the form
\be{pot}
V(r) = \left\{\begin{array}{lcl} 
    \frac Q{2R}(r^{2}/R^{2}-3)&\mbox{ for }& r\le R\\
    -\frac Qr&\mbox{for}&r>R.
  \end{array}\right.
\ee
The potential is within the CC ($r<R$) harmonic with the frequency $\Omega =
(Q/R^{3})^{1/2}$ and for $r>R$ of pure Coulomb nature.
The  full Hamiltonian of the  interacting electronic system is given by
\be{ham}
H = \sum_{j=1}^{N} \left(\frac{\vec p_{j}^{2}}2 + V( r_{j})\right)+\frac 12 \sum_{j,k=1}^{N}{\!\!\!}'  \frac 1{r_{jk}}\,,
\ee
with $ r_{jk}= \left| \vec r_j - \vec r_k \right|$ and the prime excluding the term $j=k$. 
The total ground state energy $E_{\mathrm{gs}}$ may be approximated by considering a homogeneous electron distribution. Then we can replace the sum over all electrons by an average potential energy $\vc$ and an average electron-electron interaction energy $\vee$, i.\,e.,
\be{Etot}
E_\mathrm{gs} = \sum_{j=1}^{N}V(r_j)+\frac 12\sum_{j,k=1}^{N}{\!\!\!}'\frac 1{r_{jk}}\equiv N \vc+\frac N2  \vee\,.
\ee
This assumption should be valid for sufficiently many electrons $N$, but applies also for quite small $N$ as we will see below.
Within this approximation the average energy from the background potential reads
\be{vc}
 \vc=\frac{3}{R^3} \int_0^R\!\!\mathrm{d}r\,r^2\,V(r) = -\frac 65\frac QR\,.
 \ee
The interaction energy $\vee$ cannot be calculated this way, since we have to exclude explicitly the self-interaction term. Instead we estimate the average energy of an electron $\varepsilon=\vc+\vee$ in the ground state. This energy can be approximated under the assumption that the electron density is identical to the ion density apart from a ``hole'' with volume $(4\pi/3)R^3/N$. The corresponding potential energy of such a hole is most easily calculated for a sphere (with a radius $R/N^{1/3}$) and gives
\be{Eb}
  \varepsilon =-\frac 32 \frac{N^{1/3}}{R}\,.
 \ee
This single-particle binding energy depends exclusively on the density of the CC.
Finally we can calculate the average electron-electron repulsion energy 
\be{vee}
 \vee= \varepsilon -\vc= -\frac 32\frac {N^{1/3}}R + \frac 65\frac QR\,,
 \ee
which leads to the explicit approximation
\be{Etot2}
 E_\mathrm{gs}=N\vc+\frac N2\vee = -\frac 35\frac{N^{2}}{R}-\frac 34\frac{N^{4/3}}R\,,
 \ee
for the total energy of \eq{Etot} of a neutral CC with $N=Q$.
Note, that due to the non-linear dependence of the average electron-electron interaction upon the electron density,  $E_{\mathrm{gs}}$ does not scale quadratically with $N$.  Table~1 presents  the analytically estimated energies  $\varepsilon$ and $E_{\mathrm{gs}}$ from \eqs{Eb}{Etot2}, respectively, in comparison to the numerically obtained values for 9 different values of $N$ which only for the smallest value $N=7$ differ by more than 3\,\%.

\begin{figure}
\begin{center}
\includegraphics[width=0.6\columnwidth,clip]{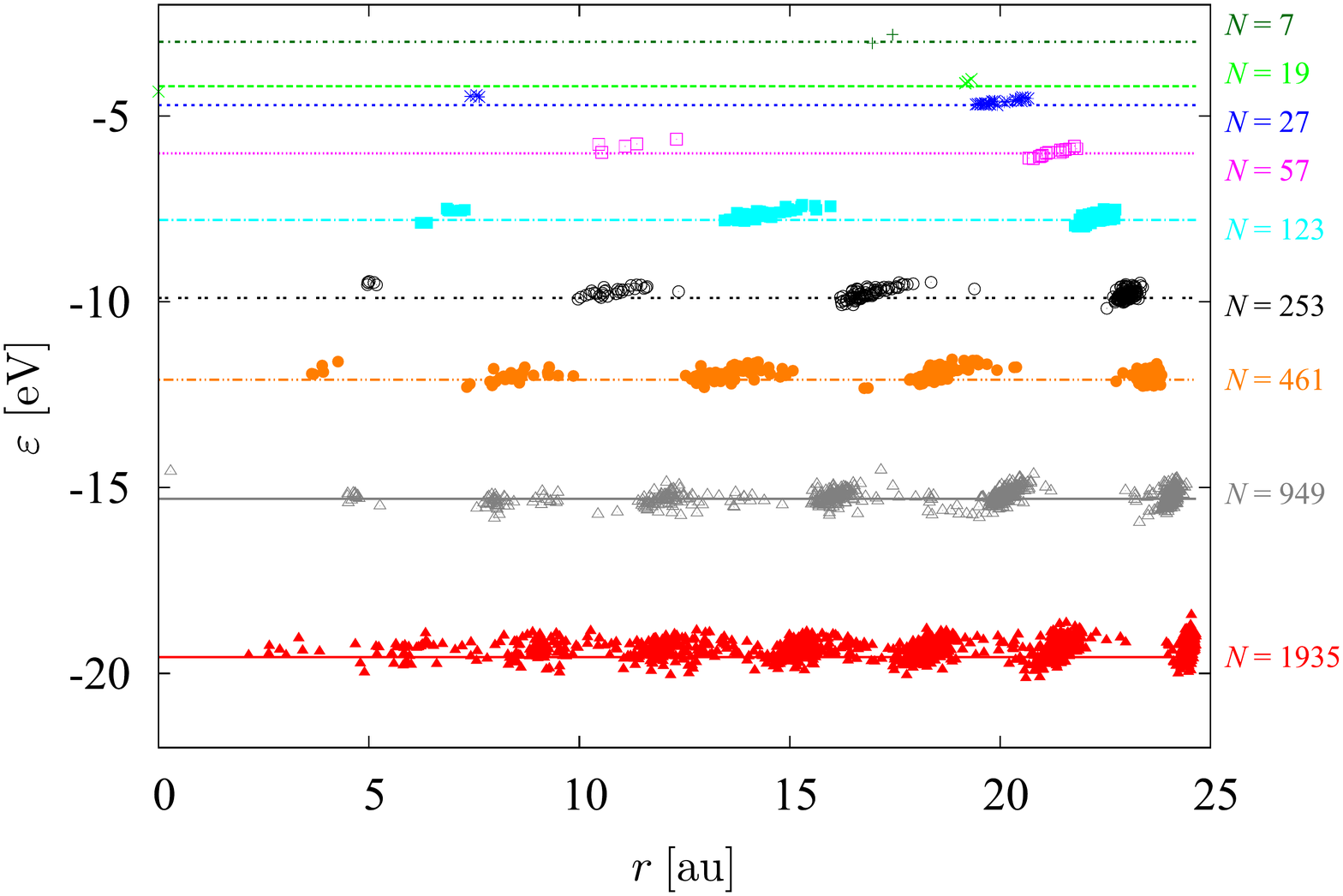}
\caption{Individual electron energies $\varepsilon_i$ according to \eq{Epsi} as a function of the electron's distance $r_i=\left| \vec r_i \right|$ from the origin of the Coulomb complex for $R=26$\,a$_0$ (Bohr radius) and various electron numbers $N$. The corresponding approximated average energies $\varepsilon$ from \eq{Eb} are shown with lines.}
\label{fig:ener}
\end{center}
\end{figure}

 The numerical energies were obtained by propagating $N$ electrons with the Hamiltonian (\ref{ham}) while reducing the particle velocities by a factor of $0.75$ every  2.4\,fs for $N \leq 123$ and every 0.24\,fs for $N > 123$ to arrive at a minimal energy configuration of the system with corresponding optimized positions $\{\bar{\vec r}_{j}\}$ for the electrons. The binding energies of the $N$ electrons are given by
\be{Epsi}
  \varepsilon_i = V(\bar{r}_i) + \sum_{k=1}^N{\!}'\frac{1}{\bar{r}_{ik}}\,.
 \ee
\Fig{fig:ener} shows these binding energies for various CCs with the same radius $R$ but different electron numbers $N$. Interestingly, not only the average values $\varepsilon$ agree quite well with the analytical estimates (shown by lines in \fig{fig:ener}), there is also relatively small variation in the individual numerical electron energies (shown by symbols in \fig{fig:ener}). This applies down to such small electron numbers as $N=7$.
In the figure one also clearly recognizes the formation of electron shells as known from so called Coulomb crystals \cite{luko+05}. 

\begin{table}[b]
  \begin{center}
    \begin{tabular}{|r|ccccccccc|}\hline
      $N$ & 7 & 19 & 27 & 57& 123 & 253 & 461 & 949 & 1935 \\\hline
      $-\varepsilon $ [eV] (num.)& 2.9& 4.1 & 4.6 & 5.9 & 7.7 & 9.8 & 12.0 & 15.3 & 19.4 \\
    (anl.)  &\it 3.0& \it 4.2&\it 4.7& \it 6.0& \it 7.8 &\it 9.9 &\it 12.1 &\it 15.3 & \it 19.6 \\
      $-E_{\mathrm{gs}}/N$ [eV] (num.) & 6.1& 14.4& 19.7& 39.4&  81.9 & 163.5 & 296.7 & 605.1 & 1226.7 \\
   (anl.)   &\it 5.9& \it 14.0&\it 19.3&\it 38.8& \it 81.1&\it 163.8 &\it 295.5 &\it 603.6 & \it 1224.9 \\ \hline
    \end{tabular}
  \end{center}
  \caption{\label{tbl:1}
    Comparison of numerical (num.) and analytical (anl.), cf.\  \eqs{Eb}{Etot2},  binding ($\varepsilon$) and ground state 
    ($E_{\mathrm{gs}}$) energies of a Coulomb complex with $N$ electrons relaxed in a background potential of \eq{pot} with $Q=N$ and radius $R = 26$\,a$_0$.}
\end{table}

\section{Photo activation of the Coulomb complex} 
\label{sec:activation}

In the previous section we have characterized the electrons of the CC in energy and prepared them in space to absorb photons.  Due to the small dipole matrix elements in the continuum it is much more likely that $\Nph$ photons of XUV energy from an intense pulse are absorbed by $\Nph$ different bound electrons compared to the situation of multi-photon absorption by a few electrons. Hence, we have the $\Nph$ photons absorbed each by one bound electron. 
The single-photon absorption rate is proportional to the intensity $I(t)$ of the pulse,
 \be{rate}
 \frac{dn(t)}{dt} = N_\mathrm{at}\sigma_{\omega} \frac{I(t)}{\omega}
 \ee
with $n(t)$ the number of electrons that have absorbed a photon up to time $t$,
while $N_ \mathrm{at}$ is the number of atoms and $\sigma_{\omega}$ the photo-absorption cross section at frequency $\omega$. 
Assuming for convenience (in fact one can use any pulse shape) a Gaussian pulse
with full width at half maximum $T$ we get 
\begin{eqnarray}
n(t) &=&\frac{N_\mathrm{at}\sigma_{\omega} I_{0}}{\omega} \int^{t}_{-\infty} dt' \exp(-4 \ln 2(t'/T)^{2}) \nonumber\\
&=&\frac {\Nph}2\left[1+ \mathrm{erf}\left( 2 \sqrt{\ln 2}\frac t
    {T} \right)\right],
\label{nphot}\end{eqnarray}
with $\Nph=N_\mathrm{at}\sigma_{\omega} I_{0} T \left( \pi / 4 \ln 2 \right)^{1/2} / \omega$ the total number of photons absorbed from the pulse.
The expression for $\Nph$ requires that each atom is singly ionized by the pulse with the same cross section $\sigma_{\omega}$. This is of course an idealization, since realistically the cross section changes even if each atom is only singly ionized through ionization into a continuum which already contains electrons. Moreover, not the highest occupied molecular orbital may get predominantly ionized leading to more than one electron per photon through auto-ionization, and finally, simply more than one photon could be absorbed by each atom leading to multiple ionization. Nevertheless, \eq{nphot} provides a reasonably general yet realistic form for the number of absorbed photons.

Within our model we account for the photo absorption by the \emph{activation\/} of electrons,
which means that only after a certain time do the electrons take part in the dynamics with an initial momentum specified below.
This implies that each electron $j$ is held fixed at its original position until its individual time $t_j$ of activation.
Since the photo absorption is a statistical process we treat the activation process statistically.
Therefore we first calculate $\Nph$ random numbers $\{x_j\}$ between 0 and 1.
The activation time $t_j$ of electron $j$ is then given by solving the equation
$n(t)=x_j\,\Nph$ which is implicit in $t$.  
In order to keep the computational expense as low as possible the number of electrons $N$ of the CC is set equal to the number of activated electrons: $N=\Nph$. For simplicity, we consider a  CC  which is neutral in the beginning, so that finally $Q = N = \Nph$.

We determine the initial momenta $\bar{\vec p}_{j}$ of the photo-activated electrons under the condition that activation of a single electron should lead asymptotically (if this electron is removed from the CC) to the atomic excess energy $\varepsilon^{*}$. This implies for the initial kinetic energy for each activated electron in the complex
 \be{Eexc}
 \frac{\bar{\vec p}^{2}_{j}}2 = \varepsilon^{*} - V(\bar{\vec r}_{j}) - \sum_{k = 1}^{N}{\!}'\frac 1{\bar{r}_{jk}}\,.
 \ee
For large $N$ we can estimate the initial average kinetic energy per electron $\varepsilon_{\mathrm{kin}}$
  by averaging \eq{Eexc} over all electrons and obtain with \eq{Eb}
  \be{kav}
 \varepsilon_{\mathrm{kin}} = \frac 1N \sum_{j=1}^{N}{\bar\vec p}^{2}_{j} = \varepsilon^{*}- (\vc + \vee) = \varepsilon^{*}+\frac 32\frac {N^{1/3}}R\,.
 \ee
 Since the numerical  ground state energies for the individual electrons are quite homogeneously distributed this also carries over to the kinetic energies $\bar{\vec p}_{j}^{2}/2$ which are close to the average values $\varepsilon_{\mathrm{kin}}$. 
Note that $\omega$ appears here only indirectly through determining $\varepsilon^*=\omega - E_{\mathrm{ip}}$
with the true atomic ionization potential $E_{\mathrm{ip}}$. 
 
Formally we can express the activation by means of an activation function\footnote{The Heaviside step function is
$\Theta(t)=\left\{\begin{array}{lcl}
    0 &\mbox{for}& t<0\\
    1 &\mbox{for}& t>0
  \end{array}\right.$}
\be{sf}
A_{\tau}(X,Y,t)=\Theta(\tau{-}t)\,X+\Theta(t{-}\tau)\,Y
\ee
in order to rewrite the original Hamiltonian (\ref{ham}) as
\begin{eqnarray}\label{hama}
H_A &=& \sum_{j=1}^{N} \left(\frac{A_{t_j}(\bar{\vec p}_j,\vec p_j,t)^{2}}2 + V\big(A_{t_j}(\bar{r}_j,r_j,t)\big)\right)\nonumber\\ &&
+\frac 12 \sum_{j,k=1}^{N}{\!\!\!}'  \frac 1{\left|A_{t_j}(\bar{\vec r}_j,\vec r_j,t)-A_{t_k}(\bar{\vec r}_k,\vec r_k,t)\right|}\,.
\end{eqnarray}
For this Hamiltonian the position of electron $j$ in phase space is kept constant at $(\bar{\vec r}_j,\bar{\vec p}_j)$ until its activation at time $t=t_j$.
One should note that, however, the single-electron energy $\varepsilon_j$ may change already for times $t<t_j$ due to the interaction with previously activated electrons. This is quite important since in highly charged systems the effects of the surrounding charges (binding, screening, etc.) easily exceed atomic properties by far and may change on a femtosecond time scale. The calculation of all interactions (including those with the not yet activated electrons) is therefore crucial. 

Note that, although the Hamiltonian $H_A$ is time-dependent through the activation functions $A$, the corresponding total energy $E(t)$ is conserved if one starts the propagation at the initial point in phase space, i.\,e., $\vec r_j(t_j)=\bar{\vec r}_j$ and $\vec p_j(t_j)=\bar{\vec p}_j$ for $j=1,\ldots,N$.
It is $\lim_{\delta t\to0}\left[E(t_j{+}\delta t)-E(t_j{-}\delta t)\right]=0$ for any $j$.

\section{Scaling in the dynamics of Coulomb complexes}\label{sec:scaling}

 The CCs including their photo activation are fully determined by a number of 
external parameters, namely $(Q,R,\varepsilon^{*},T)$. 
Before we discuss the dynamics of the CC with the initial conditions of the electrons in phase space $(\bar{\vec p}_{j},\bar{\vec r}_{j})$,
we elaborate on a universal scaling of  driven Coulomb explosion dynamics. It is based on the fact that {\it all} potentials (including the ionic background potential) originate ultimately from the homogeneous Coulomb forces. It will reveal the dynamics is the same for parameter sets emerging from 
an arbitrary reference set $(Q,R,\varepsilon^{*},T)$ by a scaling transformation which we will specify.

The system Hamiltonian \eq{ham} exhibits a global scaling  with scaled variables $\tilde x$ according to
\be{scaling}
(\vec p,\vec r) = (\eta^{1/2}\,\tilde\vec p,\eta^{-1}\,\tilde\vec r)\qquad
(E,t) = (\eta\, \tilde E, \eta^{-3/2}\,\tilde t),
\ee
which applies to Coulomb systems. Note, that the external potential \eq{pot} belongs to this class since it is given as the Coulomb potential of an extended charge distribution. 
These scaling properties can be used to identify sets of different external parameters which will lead to the same driven Coulomb explosion dynamics,
$(Q,R,\varepsilon^{*},T)\to (Q,\eta^{-1}R,\eta\varepsilon^{*},\eta^{-3/2}T)$.
E.\,g., with $\eta = (T/T^{*})^{2/3}$ we can map 
\be{mapping}
(Q,R,\varepsilon^{*},T)\to (Q,\eta^{-1}R,\eta\varepsilon^{*},T^{*}),
\ee
where $T^{*}$ is any reasonable chosen reference pulse length.
It should be emphasized, that the activation Hamiltonian $H_A$ in \eq{hama} has the same scaling properties, since the step function $\Theta$  does not posess a time scale, i.\,e.,
$\Theta(t-\tau)=\Theta\left(\eta^{3/2}(\tilde{t}-\tilde{\tau})\right)=\Theta(\tilde{t}-\tilde{\tau})$. 

\begin{figure}[htb]
\begin{center}
\includegraphics[width=0.9\columnwidth]{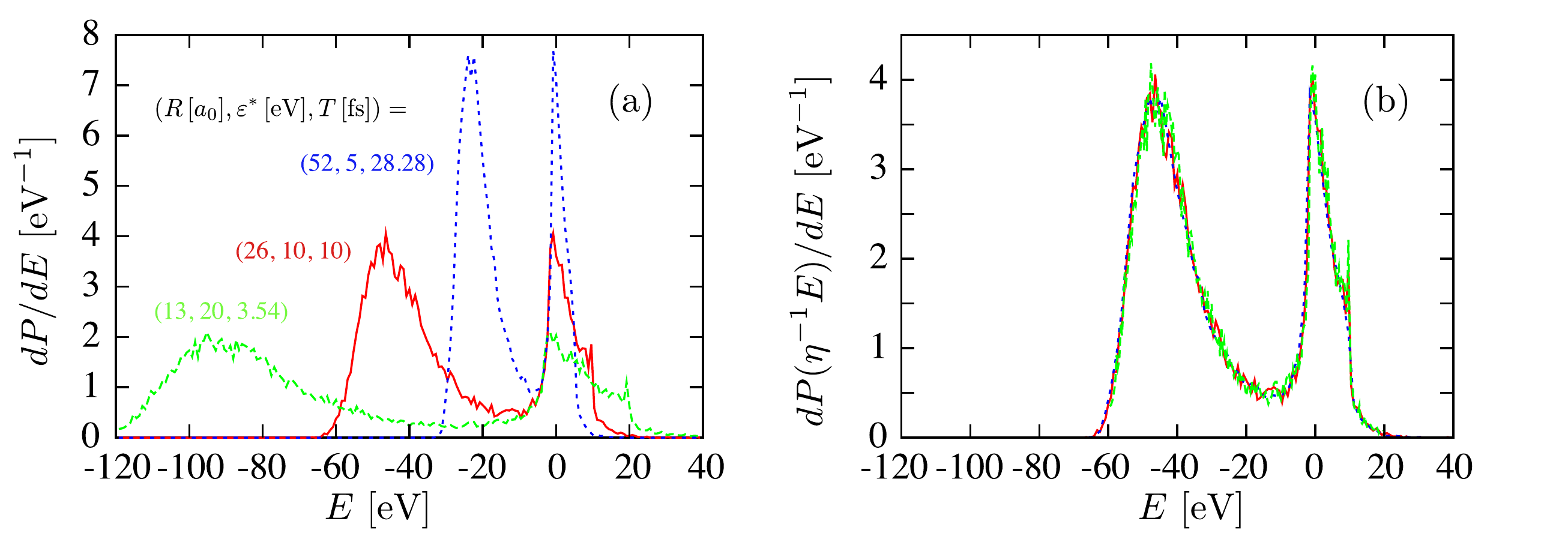}
\caption{Electron spectra $dP/dE$ for reference parameters $(Q,R\,[\mathrm{a}_0],\varepsilon^*\,[\mathrm{eV}],T\,[\mathrm{fs}])=(123,26,10,10)$ (red, full), $(123,13,20,3.54)$ (green, dashed) and $(123,52,5,28.28)$ (blue, dotted). The latter two parameters are related to the former by scaling factors of $\eta=2$ and $\eta=1/2$ respectively. Panel (a) shows the actual electron spectra $dP/dE$ and panel (b) the rescaled spectra $\eta dP(\eta^{-1}E)/dE$ that demonstrate the scaling invariance of the underlying dynamics to within numerical accuracy.}
\label{fig:scaling}
\end{center}
\end{figure}

We illustrate this scaling property in \fig{fig:scaling} where we have shown the converged electron spectra $dP/dE$ of three CCs with parameters related by the scaling property \eq{mapping}. 
In \fig{fig:scaling}a each spectrum displays the same basic features, mainly a narrow peak near $E \approx 0$ and a broad peak in the negative energy region, resulting from a complex dynamics of electron emission and equilibration of the remaining electron plasma in the activated CC which we discuss in \sec{sec:tdspectrum}. The scaling property becomes manifest in \fig{fig:scaling}b, where the spectra are rescaled according to $\eta dP(\eta^{-1} E)/dE$ to reveal that they are indeed identical within numerical accuracy.

\section{Time-resolved electron spectrum}
\label{sec:tdspectrum}
In order to understand the general feature of a double peaked final electron spectrum for an activated CC, as shown in the previous section, it is necessary to appreciate that CCs are open systems and as such are susceptible to electron loss. Indeed, it is the emission of energetic electrons that facilitates the relaxation of the remaining electrons from the highly excited activation state. This comes about as the remaining electrons in the CC lose the interaction energy with respect to emitted electrons, leading to a deeper effective binding potential.  

\begin{figure}[htb]
\begin{center}
\includegraphics[width=0.75\columnwidth]{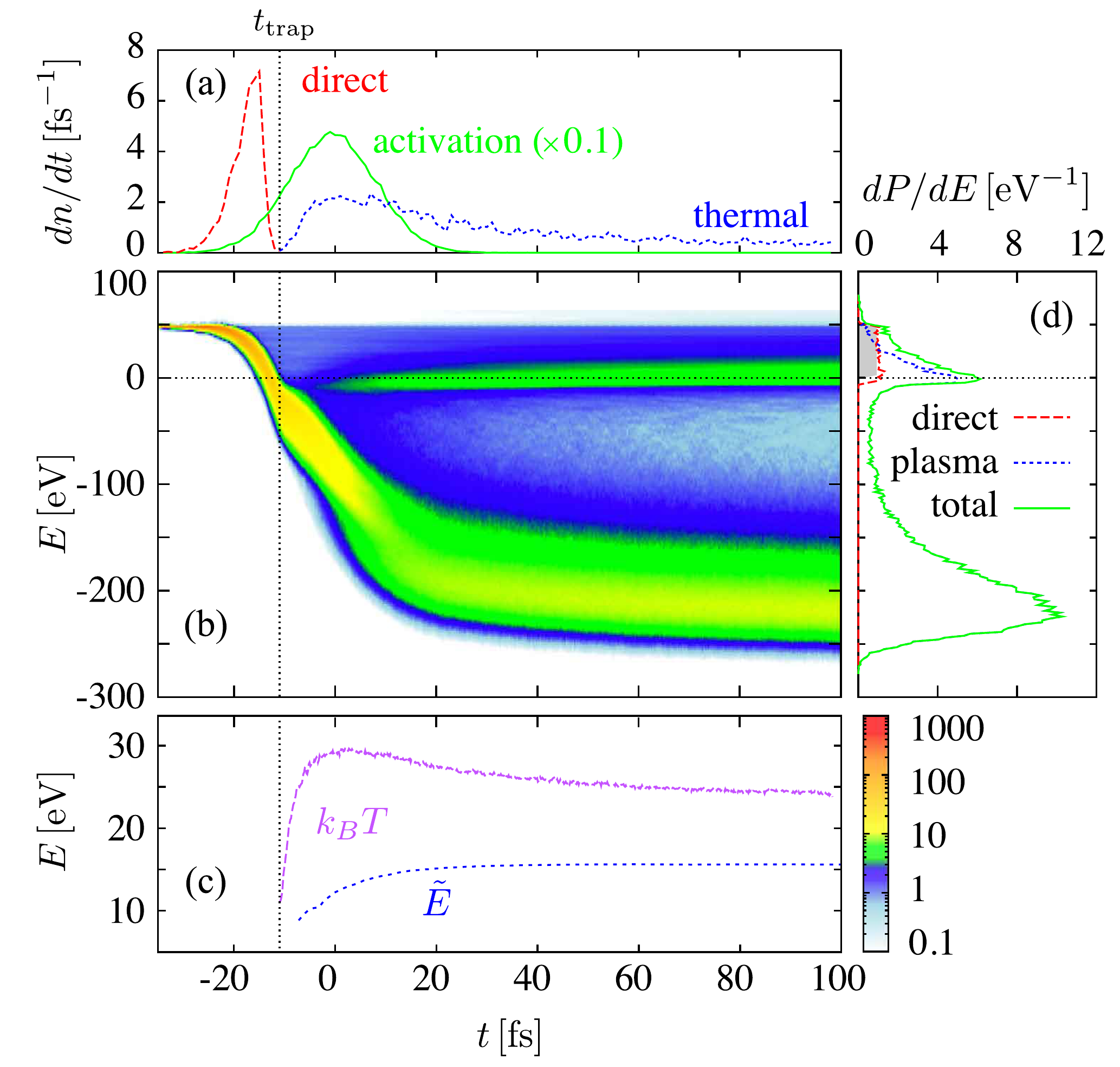}
\caption{Time evolution of an activated Coulomb complex with the following parameters $(Q,R\,[\mathrm{a}_0],\varepsilon^*\,[\mathrm{eV}],T\,[\mathrm{fs}])=(10^3,30,50,10)$. (a) Electron emission rate through direct emission before trapping time $t_{\mathrm{trap}}$ (red) and through thermal emission from a trapped plasma after $t_{\mathrm{trap}}$ (blue) and activation rate of electrons (green). (b) Time-resolved electron spectrum of Coulomb complex including as yet not activated electrons, (d) final electron spectrum at $t=100$\,fs, separated by activation time of electrons before $t_{\mathrm{trap}}$ (direct, red) and after $t_{\mathrm{trap}}$ (thermal, blue) as well as total spectrum (green). Gray shaded area marks the analytical value for a sequential spectrum (see Appendix). (c) Mean energy of the electrons in time from a Maxwell-Boltzmann fit to the kinetic energy spectrum of electrons (see (\fig{fig:mb}a)) within  a sphere of radius $2R$ (purple) and expressed through the energy scale  $\tilde{E}$ (see \fig{fig:mb}b) from the exponential fit to the spectrum of the thermally {\it emitted} electrons (blue).}
\label{bigfig}
\end{center}
\end{figure}

Insight into the formation of the final electron spectrum of a CC is obtained by tracing its time-resolved evolution from the well specified initial state described above. 
This is shown in \fig{bigfig} for parameters $(Q,R\,[\mathrm{a}_0],\varepsilon^*\,[\mathrm{eV}],T\,[\mathrm{fs}])=(1000,30,50,20)$. In this spectrum all electrons are considered, activated as well as not yet activated in which case the designated kinetic energies according to \eq{kav} are already included. Therefore the initial spectrum is mono-energetic with all electrons having a single-particle energy of $\varepsilon^*$.

\subsection*{Directly emitted electrons}
In the early stage of the activation, with a rate shown in green in \fig{bigfig}a, activated electrons have a positive energy and they may leave the CC, a process which we call {\it direct} electron emission. With each emitted electron the total charge of the CC increases thereby deepening the trapping potential for the subsequently activated electrons, which is visible as the decrease in energy of the main peak in the time-resolved electron spectrum.

The direct emission process continues until the main peak of mostly still dormant electrons falls below the threshold of $\varepsilon=0$ and newly activated electrons find themselves bound to the CC, which they can no longer leave. We refer to this as trapping of electrons \cite{saro02,halo+04,gnsa+09} by the now positively charged CC. In terms of the direct emission rate, which is shown in red in \fig{bigfig}a, this corresponds to a sharp drop-off. We call the time when the emission rate reaches zero the trapping time $t_{\mathrm{trap}}$ which here occurs near $-10\,\mathrm{fs}$.
The process of direct emission of electrons is identical to the previously described multi-step ionization \cite{both+08}. It results in a plateau-shaped spectrum shown in red in \fig{bigfig}d. The sequential process is amenable to an analytic description giving an exact expression for the height of the plateau of purely geometrical origin as $\Lambda R$, with $\Lambda \approx 0.84$ (cf.\ \ref{sec:sequential} for details).
The number of directly emitted electrons may therefore be readily approximated as $N_{\mathrm{direct}} \approx \Lambda \varepsilon^*  R$ which here amounts to less than 5\% of the total number of electrons.

\subsection*{Plasma formation and equilibration}
With the onset of trapping at $t=t_{\mathrm{trap}}$ the activated electrons within the CC undergo energy-exchanging collisions that lead to relaxation.
This has two consequences: on the one hand the spectral peak undergoes a broadening as a thermalized plasma is formed. On the other hand particularly fast electrons in the plasma may leave the complex contributing to a second emission peak in \fig{bigfig}a and resulting in a continued 
decrease of the mean energy of the plasma.
Since emitted electrons carry away energy the mean electron energy in the complex is reduced.
This opens a gap in the spectrum that individual electrons need to overcome in order to leave the cluster.
However, the plasma temperature eventually does not suffice to produce fast enough electrons and the emission decreases. From this time on the spectrum has essentially assumed its final shape. The peak at negative energy describes the thermalized plasma and the peak at positive energies contains the electrons emitted early during the activation. The gap between these two peaks must necessarily span a multiple of the plasma temperature to prevent further electron emission.

\begin{figure}[htb]
\begin{center}
\begin{tabular}{cc}
\includegraphics[width=1.0\columnwidth]{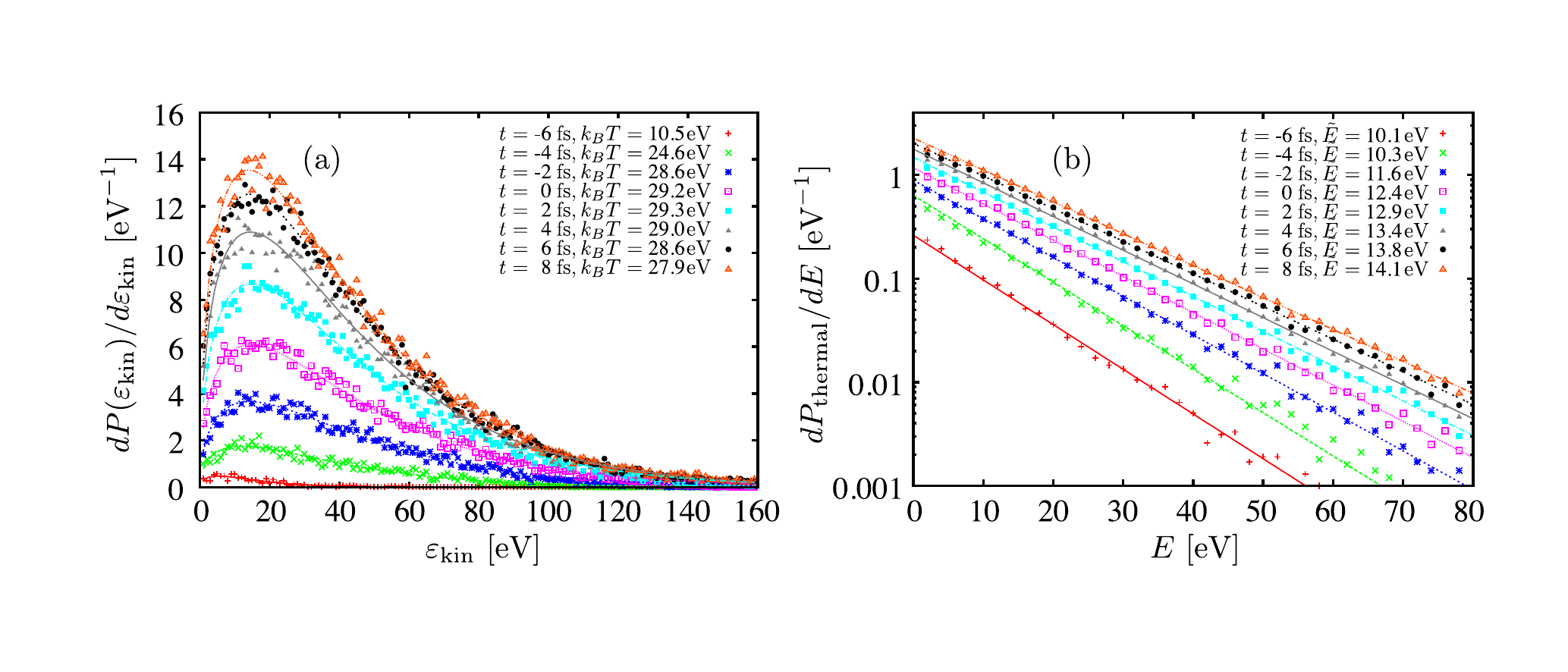}
\end{tabular}
\caption{(a) Kinetic energy spectra of plasma electrons $dP(\varepsilon_{\mathrm{kin}})/d\varepsilon_{\mathrm{kin}}$ and fitted Maxwell-Boltzmann distributions $\sim \sqrt{E/\pi k_{\mathrm{B}}T^3} \exp (-E/k_{\mathrm{B}}T)$ at different times in the evolution of the Coulomb complex. (b) Spectrum $dP_{\mathrm{thermal}}/dE$ of thermally emitted electrons at various times and exponential fit $\sim\exp(-E/\tilde{E})$.}
\label{fig:mb}
\end{center}
\end{figure}%

The kinetic energy distribution  of the plasma electrons follows closely a Maxwell-Boltzmann distribution, as can be seen  in \fig{fig:mb}a. Since the photo activation continues at the time for which the distributions are determined, we consider only electrons activated before  $t-t_{\Omega}$, where $t_\Omega=2\pi (R^3/Q)^{1/2}$ is the oscillation period within the Jellium potential which provides an intrinsic time-scale of the electron dynamics. Obviously, this time is sufficient for the equilibration of newly activated electrons.

The plasma temperature extracted from the Maxwell-Boltzmann fit  exhibits a sharp rise beginning at time $t_{\mathrm{trap}}$ towards a maximum close to the time of maximum activation rate (purple, \fig{bigfig}c).   
The rise reflects the increasing depth of the positive background potential allowing to trap  faster electrons in a hotter plasma.
Eventually, the temperature decreases due to evaporative cooling effects and levels off asymptotically slightly below 25\,eV. 
Note, that this temperature remains consistently below the temperature inferred from the initial kinetic energy of the electrons, \eq{kav}, which is due to a broadened spatial distribution of the plasma with respect to the ground state. 

\subsection*{Emission from the plasma}
As can be seen in \fig{fig:mb}b and as has recently been similarly described elsewhere \cite{zila+09,arfe10}, the spectrum  $dP_{\mathrm{thermal}}/dE$  of the electrons emitted by the thermalized plasma forms an exponential distribution characterized by the energy scale $\tilde{E}$,
\be{emitted}
 dP_{\mathrm{thermal}}/dE  = N_{\mathrm{thermal}} \tilde{E}^{-1} \exp (-E/\tilde{E})\,,
 \ee 
where $N_{\mathrm{thermal}}$ is the number of electrons thermally emitted from the plasma.
Although  originating from the tail of the equilibrated plasma inside the cluster the ``temperature''
 $\tilde E$ of the thermally emitted electrons (blue, \fig{bigfig}c)  is considerably lower than $k_{B}T$ of the trapped plasma (purple, \fig{bigfig}c). One should therefore not take this experimentally directly accessible energy $\tilde E$ as the temperature of the trapped electron plasma. 

The exponential distribution \eq{emitted} has been measured recently \cite{both+10} in an experiment, where  ionization of clusters with  90\,eV photons from intense pulses supplied by FLASH \cite{acas+07} was studied by electron spectroscopy.
The electron spectra of irradiated xenon clusters exhibit high-energy tails of the photo lines, which
 extend further with increasing intensity. 
These blue wings in the spectra can be related to  the formation of an electron  plasma with supra-atomic density, generated due to the large photo-absorption cross section of xenon at 90\,eV.
Theoretically, these electrons could be distinguished from those emitted directly at the beginning of the pulse \cite{both+10}. Note that electron emission may also occur for photo activation just below threshold, $\varepsilon^*\,{\stackrel{<}{{\mbox{{\tiny$\sim$}}}}}\,0$, as measured recently at SCSS \cite{shta+10} for neon clusters \cite{ue10}. Whereas direct emission is not possible, the energy of the excited electrons is redistributed through collisions with some electrons being emitted.

\section{Summary}

We have discussed a simple and fairly general scenario for the electron dynamics in dense matter exposed to short  and intense laser pulses of high intensity. It applies to various frequencies $\omega$ of the light up to X-rays. The lower limit for $\omega$  is set by the condition that 
single-photon ionization of many atoms dominates which we call photo activation. We have provided a simple formalization of the photo activation in time which relates the successive photo-ionization events to the time-dependent shape of the laser pulse and the total number $N$ of ionized electrons generated. The process
leads inevitably to a Coulomb complex, a many particle system of ions and electrons whose Coulomb interaction dominates all other forces. From this follows a simple scaling property of Coulomb systems which relates seemingly disparate experimental szenarios with respect to the size of the system (characterized by the radius $R$ of atomic positions), the time scale of photo activation (determined by the pulse width $T$) and the electron  energy $N(\omega -E_{\mathrm{ip}})$ made available through the activation. The latter depends through $N$ on the  total energy contained in the laser pulse.

Furthermore, we have discussed the time-evolution of the electrons during and after  photo activation including the formation of a plasma and its equilibration in detail. Photo activation in Coulomb complexes is not limited to the examples of intense XUV and X-ray pulses illuminating clusters of almost solid density as discussed here. 
The very same phenomenon occurs, e.g., in the activation of ultracold plasmas \cite{kipa+07}  from atom clouds held in magneto-optical traps at appropriately adopted time and energy scales.

\ack

The authors thank Ionu\c{t} Georgescu for helpful discussions.
US and JMR acknowledge support by the KITP at UC Santa Barbara during the program {\it X-ray Frontiers}.

\appendix 
\section{Electron spectrum for sequential ionization}
\label{sec:sequential}
We derive an analytical expression for the electron spectrum in the case of high excess energy $\varepsilon^* \gg Q/R$ and long pulses $T \rightarrow \infty$. In these limits all electrons leave the CC directly upon activation, without the possibility of exchanging energy with other active electrons, i.\,e., the emission occurs sequentially.
Furthermore we assume a homogeneous electron distribution within $r<R$ of the non-activated electrons, which gives rise to the probability distribution in the radial coordinate for the electron to be activated next as
\begin{eqnarray}
\frac{dP}{dr} = \frac{3r^2}{R^3}.
\end{eqnarray}
\begin{figure}[b]
\begin{center}
\includegraphics[width=0.5\columnwidth]{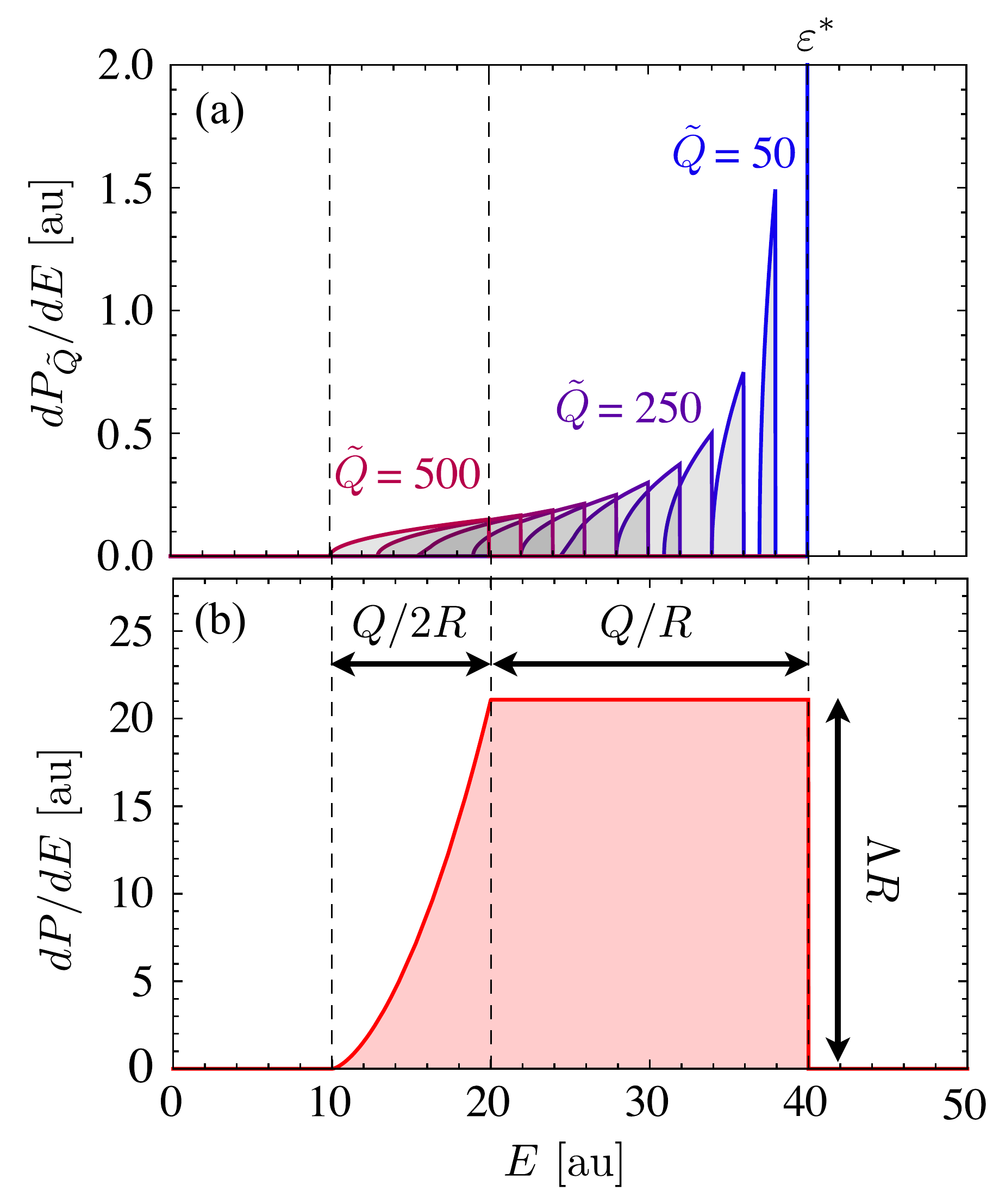}
\caption{(a) Probability distribution in energy $dP_{\tilde{Q}}/dE$ from \eq{eq:pdoe} of next activated electron for CC with parameters  $(Q,R\,[\mathrm{a}_0],\varepsilon^* \,[\mathrm{au}])=(500,25,40)$ and charge $\tilde{Q}=0,50,\dots,500$ on the CC. (b) Analytical electron spectrum $dP/dE$ from \eq{aspec} from integration of probability distributions.} 
\label{fig:seq}
\end{center}
\end{figure}%
Released electrons leave a charged CC behind.
This remaining charge, denoted by $\tilde{Q}$, is within our approximation also homogeneously distributed.
It modifies the asymptotic (or measured) energy $E$ of the next to be activated electron.
Therefore the energy depends on the radial coordinate $r$, 
\be{etilde}
E_{\tilde{Q}}(r)=\varepsilon^* -\frac{3}{2} \frac{\tilde{Q}}{R}+\frac{\tilde{Q}}{2 R^3}r^2,
\ee
with the modification given by the potential in \eq{pot}.
The dependence on the instantaneous charge $\tilde{Q}$ is emphasized by the subscript.
This expression allows one to obtain the probability distribution in terms of the energy as
\begin{eqnarray}
\frac{dP_{\tilde{Q}}}{dE} &=& \frac{dP}{dr}\bigg/ \frac{dE_{\tilde{Q}}}{dr} 
\nonumber\\
&=& \frac{3}{\tilde{Q}}r
=\frac{3}{\tilde{Q}}\left(\left( E-\varepsilon^*+\frac{3}{2} \frac{\tilde{Q}}{R} \right)
 \frac{2 R^3}{\tilde{Q}} \right)^{1/2}, \label{eq:pdoe}
\end{eqnarray}
where the lower line uses the inverse function of \eq{etilde}.
As the radial coordinate is restricted to $0\le r\le R$, so is the energy in \eq{eq:pdoe}, namely
${-}(3/2)\tilde{Q}/R \leq E{-}\varepsilon^* \leq  {-} \tilde{Q}/R$.
The reduction of the energy $E{-}\varepsilon^*$ is bounded by the potential energy at the centre and the surface. 
The probability distribution of \eq{eq:pdoe} is shown for the specific case of $\varepsilon^*=40\,\mathrm{au}$, $R=25\,\mathrm{a}_0$ und various instantaneous charges $\tilde{Q}=0,50,\dots,500$ in \fig{fig:seq}a. 
The larger the charge on the CC is, the wider the distribution in energy becomes and the further it moves to lower energy. 

\begin{figure}[b]
\begin{center}
\includegraphics[width=0.5\columnwidth]{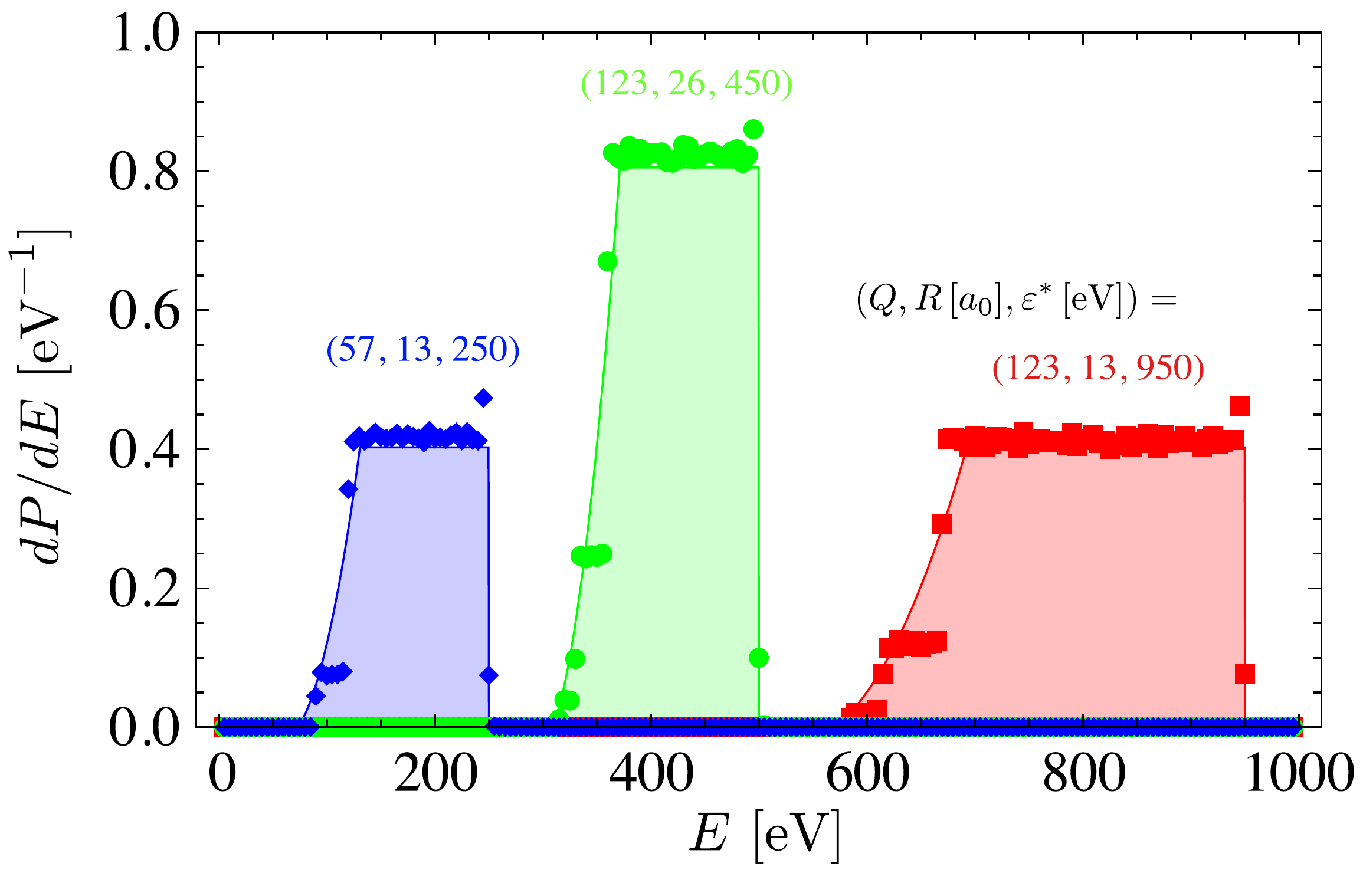}
\caption{Comparison of numerical (symbols) and analytical (lines and shaded) electron spectra for various parameters. Analytical spectra calculated according to \eq{aspec}.  A pulse length of $T=10$\,fs, sufficiently long for sequential ionization, was used for all three spectra.} 
\label{fig:seq2}
\end{center}
\end{figure}%
As can be seen in \fig{fig:seq} only certain values of $\tilde{Q}$ contribute to  the spectrum at a given energy $E$. 
These values are given by $\tilde{Q}_{1}(E)\le \tilde{Q}\le \tilde{Q}_{2}(E)$ with
$\tilde{Q}_{1}(E)=(2/3)R(\varepsilon^*{-}E)$
and $\tilde{Q}_{2}(E)=\min \left(R(\varepsilon^*{-}E),Q\right)$. 
Remember that $Q$ is the total charge reached after the activation and removal of all electrons.
Thus one obtains by integration
\begin{eqnarray}
\frac{dP}{dE} &=& \int_{\tilde{Q}_{1}(E)}^{\tilde{Q}_{2}(E)}d\tilde{Q}\:\frac{dP_{\tilde{Q}}}{dE}
\nonumber\\
&=&\left\{\begin{array}{lll}
\Lambda\:R  &\mbox{for}&   {-}Q/R\le E{-}\varepsilon^*\le 0, \\
\lambda_{Q,R}(E)  &\mbox{for}&  {-}(3/2)Q/R\le E{-}\varepsilon^*\le {-}Q/R.
\end{array}\right.\label{aspec}
\end{eqnarray}
Interestingly the electron spectrum for higher energies is independent of energy $E$ with a system-independent constant
given by 
$\Lambda=3 \left[2\sqrt{3}\ln\left(3{+}\sqrt{3}\right){-}\sqrt{3}\ln\left(6\right){-}2\right]\approx0.843$.
Towards lower energies $E$ it falls off monotonically according to 
$\lambda_{Q,R}(E)=3R\left[2\sqrt{3}\ln\left(3{+}\sqrt{9{-}6\chi}\right){-}\sqrt{3}\ln\left(6\chi\right){-}2\sqrt{3{-}2\chi}\right]$ with $\chi=(\varepsilon^*{-}E)R/Q$. 

The entire analytical spectrum is shown in \fig{fig:seq}b for the specific parameters $Q=500$, $R=25\,\mathrm{a}_0$ und $\varepsilon^*=40\,\mathrm{au}$.
In particular the plateau in the region from $\varepsilon^*-Q/R$ to $\varepsilon^*$ is a characteristic feature of the sequential ionization as seen in an experiment \cite{both+08} and numerical simulations \cite{mo09}. While the height of the spectrum is solely determined by the radius $R$, the width of the full spectrum and of the plateau-region are given by $3Q/2R$ and $Q/R$ respectively. Note that this implies that in the case of $\varepsilon^* < 3Q/2R$ trapping of electrons will occur and the full sequential spectrum is not realized. However, for those electrons emitted before the onset of trapping the sequential spectrum as derived above is valid. 

We compare the analytical spectrum of \eq{aspec} with numerical results for three sets of parameters in \fig{fig:seq2}, in each case with a pulse length of $T=10$\,fs which proves sufficiently long for the CC to exhibit sequential ionization behaviour. Indeed, we find an overall excellent agreement between the numerical and the analytical result, in particular as pertains to the appearance of an extended plateau region, well described in width, height and position by \eq{aspec}. 
Nevertheless, some minor discrepancies owing to the continuous approach employed in the analytical calculation can be observed. A narrow peak at $\varepsilon^*$ in the numerical spectrum arises from the first activated electron with exactly the excess energy $\varepsilon^*$. Furthermore, in the energy domain ${-} 3Q/2R \leq E{-}\varepsilon^* \leq {-} Q/R$, the shell-like structure of the initial configuration of the CC, as shown in \fig{fig:ener}, leads to a step-like spectrum, as opposed to the smooth increase in the analytical case.


\section*{References}


\begin{thebibliography}{10}

\bibitem{acas+07}
W. Ackermann {\it et~al.}, Nat. Photon. {\bf 1},  336  (2007).

\bibitem{shta+10}
T. Shintake {\it et~al.}, Nat. Photon. {\bf 2},  555  (2010).

\bibitem{emak+10}
P. Emma {\it et~al.}, Nat. Photon. {\bf 4},  641  (2010).

\bibitem{fuli+09}
H. Fukuzawa {\it et~al.}, Phys. Rev. A {\bf 79},  031201 (R)  (2009).

\bibitem{both+10}
C. Bostedt {\it et~al.}, New J, Phys. {\bf 12},  083004  (2010).

\bibitem{newo+00}
R. Neutze, R. Wouts, D. van~der Spoel, E. Weckert, and J. Hajdu, Nature {\bf
  406},  752  (2000).

\bibitem{br93}
M. Brack, Rev. Mod. Phys. {\bf 65},  677  (1993).

\bibitem{luko+05}
P. Ludwig, S. Kosse, and M. Bonitz, Phys. Rev. E {\bf 71},  046403  (2005).

\bibitem{saro02}
U. Saalmann and J.~M. Rost, Phys. Rev. Lett. {\bf 89},  143401  (2002).

\bibitem{halo+04}
S.~P. Hau-Riege, R.~A. London, and A. Sz{\"o}ke, Phys. Rev. E {\bf 69},  051906
   (2004).

\bibitem{gnsa+09}
C. Gnodtke, U. Saalmann, and J.~M. Rost, Phys. Rev. A {\bf 79},  041201\,(R)
  (2009).

\bibitem{both+08}
C. Bostedt {\it et~al.}, Phys. Rev. Lett. {\bf 100},  133401  (2008).

\bibitem{zila+09}
B. Ziaja {\it et~al.}, New J, Phys. {\bf 11},  103012  (2009).

\bibitem{arfe10}
M. Arbeiter and T. Fennel, Phys. Rev. A {\bf 82},  013201  (2010).

\bibitem{ue10}
K. Ueda, private communication  (2010).

\bibitem{kipa+07}
T.~C. Killian, T. Pattard, T. Pohl, and J.~M. Rost, Phys. Rep. {\bf 449},  77
  (2007).

\bibitem{mo09}
K. Moribayashi, Phys. Rev. A {\bf 80},  025403  (2009).

\end{thebibliography}
\end{document}